# Estimation of Building Rooftop Temperature from High Spatial Resolution Aerial Thermal Images

Atma Bharathi Mani*, Ramanathan Sugumaran**

*Abstract*— This letter presents a novel technique to calculate temperatures of building rooftops and other impervious surfaces from high spatial resolution aerial thermal images. In this study, we collected aerial radiance images of 30cm spatial resolution using a FLIR Phoenix imager in long-wave and mid-wave infrared wavelengths for the city of Cedar Falls, USA to estimate building roof temperature loss. Simultaneous ground temperature measurements were made at pre-selected ground targets and roofs using 9 Fluke 561r infrared thermometers. Atmospheric correction of aerial images was performed by Empirical Line Calibration (ELC) method. The resulting ground-leaving radiances were corrected for emissivity of different roof types and true kinetic temperature of the building roofs was calculated. The ELC model was observed to perform better when only impervious surface targets were used for the regression. With an R2 = 0.71 for ELC, the method produced a root mean squared error of 0.74°C for asphalt roofs. Further, we observed that the microclimate plays a significant role while synchronizing aerial and ground measurements.

## I. Introduction

Heat loss detection from residential and commercial buildings is an important aspect of infrastructure maintenance and energy efficiency [1]. High resolution aerial thermal images make remote detection of attic insulation leaks and calculation of rooftop temperature possible [1-3]. In addition to heat loss studies, remotely computed roof-top temperatures have been used as an input in several urban heat island studies [3-10]. The coarse spatial resolution of space-borne thermal images and the non-availability of less expensive aerial thermal images had been limiting factors in applying thermal remote sensing for high resolution urban studies [4]. Further the accuracy of the temperature calculated depends on atmospheric effects namely absorption, re-emission and emissivity and geometry of the roofs [2], [3], [11] making comprehensive atmospheric and emissivity corrections a requirement.

Earlier studies involved simpler models that did a direct conversion from image digital numbers (DN) to surface temperature [12] [13] without correcting for atmospheric or emissivity variations. Studies that performed corrections for atmospheric effects, accomplished that using three major methods – radiative transfer modeling, Split Window Techniques (SWT) and using empirical models. Radiative transfer models such as LOWTRAN, MODTRAN are physics based models that require extensive and often prohibitive environmental data [14], [15]. SWT are employed when radiance data in more than one thermal band or view angles is available [16], [17]. Empirical models are created typically where simultaneous ground and aerial thermal data of known targets are available [18]. Empirical Line Calibration (ELC) is a quick and efficient empirical method that has been widely used to convert optical images in radiance to absolute reflectance [14], [19]. Different implementations of ELC exist however the basic principle is to create a statistical relationship between ground and aerial radiances of known targets and use it to predict the ground-leaving radiance at other pixels. By assuming a linear relationship between the two variables modeled, uniform atmospheric effects throughout the image and flat topography, ELC corrects the atmospheric path radiance and attenuation effects [14], [15], [20]. To accurately model the scene, it is necessary to have at least two targets with high and low radiance such that rest of the pixels would range between them [14], [20], [21]. Studies have shown that the accuracy of the statistical model and hence the method improve with increasing number of targets [1], [15]. Further the accuracy of the ELC model depends on the accuracy of field measurements and the characterization of the targets [22].

Corrections for emissivity have been performed under various levels of complexity. The spatial resolution of the thermal images was observed to influence the accuracy of emissivity correction required. Studies that did not perform any correction assumed the surface area as a blackbody or preferred to deal with brightness temperature[23], some studies applied the average emissivity of all the materials in the study area throughout the scene [12], [24] and some performed classification based emissivity, applying a class specific emissivity for different polygons in the study area [25]. Given that emissivity of a material is a function of

This work was supported in part by the U.S. Department of Energy, Office of Energy Independence.

Atma B. Mani*, graduate student, University of Northern Iowa, Cedar Falls, IA 50613 USA (e-mail: atma@ uni.edu).

Ramanathan Sugumaran**, professor, University of Northern Iowa, Cedar Falls, IA 50613 USA. (e-mail: sugu@uni.edu).

2wavelength, temperature and look angle [26], there is little consensus on an accurate source for emissivity.

Despite being efficient, ELC has been rarely used to correct thermal images mainly due to the elaborate field work and the difficulty of resolving ground targets in coarse spatial resolution images from space-borne platforms. However aerial platforms permit detailed field planning and the very high spatial resolutions (in the order of 30 cm or lesser) permit the ground data collection required for methods like ELC. In this paper, we present a comprehensive methodology to calibrate field IR thermometers, compute high precision band emissivity of targets and roof materials from spectral reflectance curves, correct for atmospheric effects using an empirical model and inter convert between radiance and temperature. The rooftop temperatures extracted by this model is intended to be used by Cedar Falls Utilities, the energy company for the city to identify buildings with poor insulation and to compare them on a quantitative basis.

## II. Methodology

In the methodology section we explain a) our ground and aerial data collection techniques, instruments used and the environmental conditions, b) calculating band emissivity by integrating spectral reflectance curves of different roof materials, c) converting ground temperature data into radiance, correcting for emissivity of target surfaces and normalizing for instrumental differences, d) regressing ground and aerial radiance as part of ELC and analyzing the results to refine the model, applying the correction to obtain ground-leaving radiance at all pixels, e) finally correcting for emissivity of different rooftops and converting the ground-leaving radiance to kinetic temperature of rooftops.

### A. Aerial and Ground Data Collection

30cm spatial resolution radiance images were collected using FLIR Phoenix LWIR (8-9.2µm) and SWIR (3-5µm) camera mounted on a fixed wing aircraft. The images were collected from 10 pm to 3 am on 18th and 19th of November 2010. On the ground, 4 teams for roofs, 7 teams for gray targets, 4 teams for black boards and water body targets were dispatched to collect simultaneous ground temperature measurements. The ground teams used Fluke 561r infrared thermometer operating in LWIR (8-14µm) region with fixed emissivity of 0.95 to record radiant temperatures of targets and a GPS unit to record the locations. Temperature from 25 building roofs of different roof materials and of different dwelling types were collected by the roof teams. The gray targets for ground teams were composed of side-walks made of concrete, roads made of asphalt or concrete and grass and 134 such targets were collected. The geographic distribution of the targets was such that each half of a flight-line had one target and the temporal distribution was such that each team directly collects the target temperature a few minutes after the aircraft had imaged that pixel. Finally 4 wooden boards (8 by 4 feet in dimension) painted with multiple coats of black matte paint were placed next to 4 large water bodies. Temperature of the boards and the water body was collected every half hour synchronizing with the university weather station. The temperature from water bodies and black boards acted as hot and cold targets and the temperature from concrete sidewalks, asphalt roads and grass acted as intermediate temperature targets. The readings from roofs were used for validation of the ELC model. However for this paper, a subset of the city entirely from day 1 images was created to perform the atmospheric and emissivity corrections. The pilot study area was about 9.3 sq. km in area with about 1400 buildings. 40 grey targets and 22 roof targets composed the ground data.

### B. Emissivity Calculation

The major land cover types relevant to this project were cement, asphalt, water (lakes, streams), grass, black matte paint (for the black boards) and metal roofs. By Kirchoff's law for opaque surfaces, absorptance α equals emittance ε at a given temperature [26].

$$\alpha = \epsilon = 1 - \rho \qquad (1)$$

The spectral emissivity can be represented as

$$\epsilon(\lambda) = \frac{R'(\lambda, T)}{R(\lambda, T)} \qquad (2)$$

where R'(λ,T) is the spectral radiant emittance of an object at a particular wavelength and temperature; and R(λ,T) is that of a black body at the same wavelength and temperature. Using equation 1, the spectral emissivity of land cover materials were calculated from their reflectance spectra taken from ASTER JHU spectral library. In practice, thermal sensors work on a broad band of wavelengths such as MWIR, LWIR and to obtain the mean emissivity in that band, equation 2 has to be integrated over the limits of the wavelengths [26]. Thus

$$\varepsilon_{\lambda_2 - \lambda_1} = \int_{\lambda_1}^{\lambda_2} \frac{R'(\lambda, T)}{R(\lambda, T)} \, d\lambda \qquad (3)$$

Further, from (2), R'(λ,T) the spectral radiant emittance of an object can be expressed as

$$R'(\lambda, T) = \epsilon(\lambda) R(\lambda, T) \qquad (4)$$

Substituting (4) in (3), we get

$$\varepsilon_{\lambda_2 - \lambda_1} = \int_{\lambda_1}^{\lambda_2} \frac{\epsilon(\lambda) R(\lambda, T)}{R(\lambda, T)} \, d\lambda \qquad (5)$$

The spectral radiance emittance of a black body at a particular temperature is obtained from fundamental Planck's law given by (6) [26].

$$R(\lambda, T) = \frac{2\pi h c^2}{\lambda^5} \frac{1}{e^{ch/\lambda k T} - 1} \qquad (6)$$

Thus in this study, for each of the land covers, the mean emissivity in 8-9.2 µm and in 8-14µm was calculated using (5). The samples used in the ASTER JHU were assumed to be at room temperature; hence 23°C (or 300K) was used as the temperature in (6). Table 1 shows the emissivity calculated for major land cover types used in this project.

TABLE I

| Material | LWIR 8-9.2 | LWIR 8-14 |
|---|---|---|
| Asphalt | 0.93084419 | 0.946427167 |
| Concrete | 0.95352648 | 0.970666471 |
| Tap Water | 0.98378147 | 0.983423632 |
| Black board | 0.95584074 | 0.948549302 |
| Grass | 0.98178366 | 0.983275663 |
| Tar | 0.95830696 | 0.958712242 |
| Tree | 0.97863716 | 0.977182788 |
| Soil | 0.91176422 | 0.955334553 |
| Roofing rubber white | 0.96455994 | 0.967303921 |
| Roofing rubber black | 0.91196318 | 0.913549303 |
| Metal roof | 0.632978 | 0.619098699 |
| Metal roof rusted | 0.72807754 | 0.790403235 |

Band emissivities of targets and rooftops found in this study. Band emissivities were computed through numerical integration of spectral emissivities obtained from spectral reflectance curves of these materials.

### C. Ground Instrument and Data Normalization

The field temperatures were collected using 9 infrared thermometers. To avoid instrument induced variation, the readings had to be normalized to a control. For this study, a beaker with distilled water was used as the medium and its temperature read from an alcohol thermometer as control. The distilled water was cooled to 3°C and allowed to rise to room temperature. The water was stirred constantly using a magnetic stirrer apparatus and the temperature from each IR thermometer was read successively for every 0.5°C rise. Temperature of the distilled water from the alcohol thermometer flanked the beginning and end of each reading set to account for temperature drift while making one cycle of measurements. The alcohol thermometer read the kinetic temperature whereas IR thermometers read the radiant temperature; hence emissivity correction on the IR thermometer reading had to be performed prior to normalization.

Let $e_0 = 0.95$ the emissivity setting on the IR thermometer, $e_1 = 0.9838$ the emissivity of distilled water calculated using (5), T be the temperature from IR thermometer for emissivity = $e_0$ setting. By integrating Planck's law between 8 to 14 µm, the radiant emittance from the distilled water as measured by the IR thermometer ($R_{e_0}$) can be found. Thus

$$R_{e_0} = \int_8^{14} R(\lambda, T) d\lambda \qquad (7)$$

However the emissivity of distilled water is $e_1$, hence the true radiant emittance from distilled water is calculated as $R_{e_1}$

$$R_{e_1} = R_{e_0} * \left(\frac{e_0}{e_1}\right) \qquad (8)$$

The radiant emittance from a black body at the same temperature and wavelength band is calculated as $R_{e_2}$ where $e_2 = 1.0$

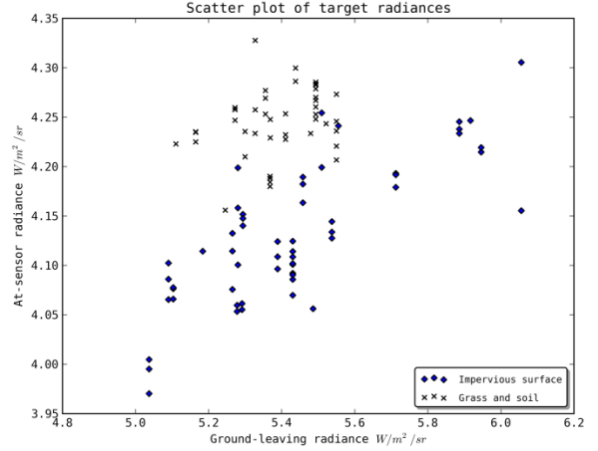

Fig. 1. Scatter plot of ground-leaving radiance vs. at-sensor radiance for all ground targets. Impervious targets show a linear trend while radiances from grass and soil targets show poor correlation.

$$R_{e_2} = R_{e_1} * \left(\frac{e_2}{e_1}\right) \qquad (9)$$

Now using the radiant emittance from a black body at same temperature, the kinetic temperature ($T_{k_{e_2}}$) of distilled water can be found by inverting Planck's law.

Finally the kinetic temperature calculated for each IR thermometer is regressed with the reading from the alcohol thermometer. The slope and offset obtained for each IR thermometer represent the normalization parameters. The IR thermometers demonstrated a strong linear trend achieving an $R_2$ greater than 0.95 in each case. Using these parameters, the ground temperatures measured were normalized for each instrument. During this process, for each measurement the appropriate emissivity of the land cover was substituted in the place of $e_1$ and the kinetic temperature of the surface was calculated.

### D. Atmospheric Correction using ELC Method

The normalized ground target temperatures were converted to radiances in 8-9.2 µm region using (7) and (8). A table with ground-leaving radiance and the corresponding aerial at-sensor radiance for the 40 non-roof ground targets was created. From Figure 1 that shows a scatter plot of ground-leaving radiance vs. at-sensor radiances, grass and soil targets can be found to be poorly correlated when compared to impervious surface targets.

Similar experience was reported by *Voogt et. al.* [5] and they were eliminated from further processing. A simple linear regression between ground and at-sensor radiance of impervious targets was performed such that it predicts the ground-leaving radiance for any at-sensor radiance. The model achieved an $R_2$ of 0.60 representing a fair linear relationship as shown in Figure 2. From the fitted vs. residual plot and Cook's distance plots, observations 4, 15 and 25 were concluded as damaging points [27]. The regression was repeated omitting these observations and the model improved to an $R_2$ of 0.71. The result of the ELC based atmospheric correction was of the form

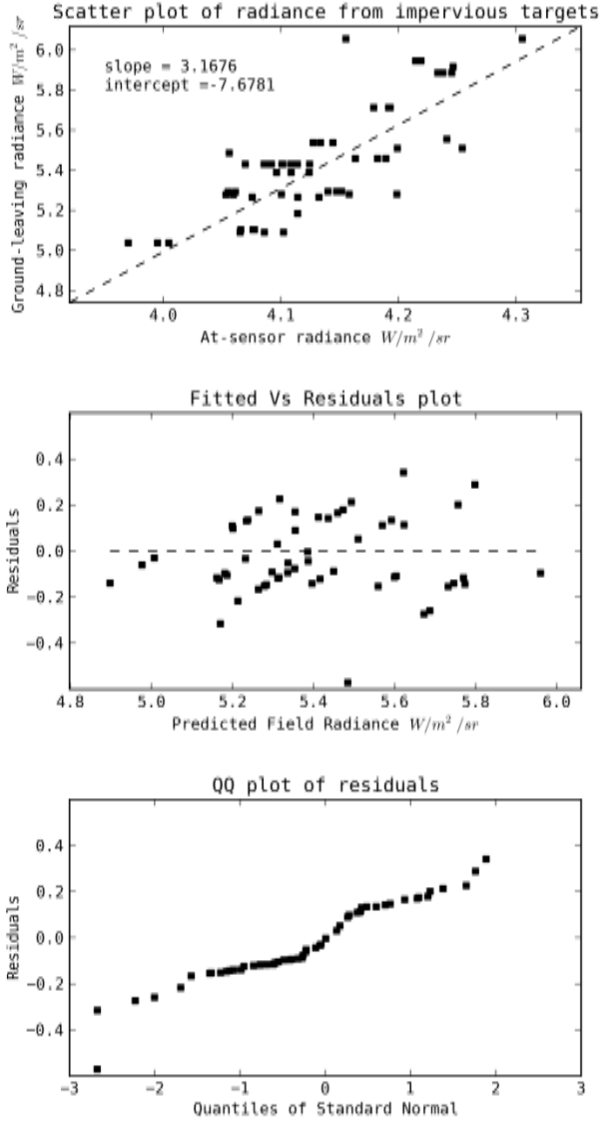

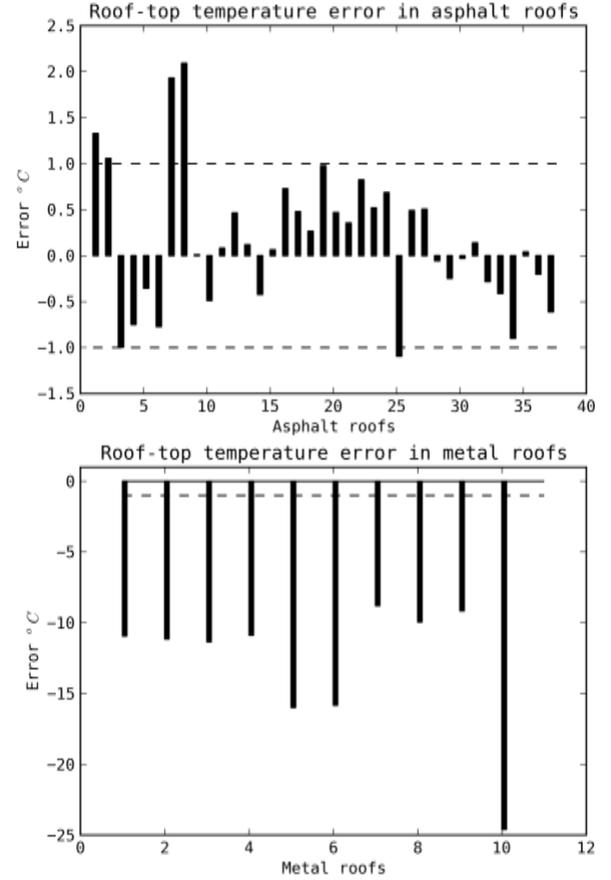

Fig. 2. Regression model evaluation for Empirical Line Calibration method. The scatter plot of at-sensor vs. ground-leaving radiance of impervious targets shows a strong linear trend. The fitted vs. residuals plot shows random distribution of errors and the scatter of quantiles of a standard normal distribution with the residuals is a fairly straight line depicting normal distribution of residuals.

$$Ground\ leaving\ radiance = 3.1676 * \\ (sensor\ reaching\ radiance) - 7.6481 \quad (10)$$

### E. Emissivity Correction

Using (10), the sensor reaching radiances from all aerial images were atmospherically corrected and converted to ground-leaving radiance. Building roof materials in the study area was analyzed and reduced to 4 dominant types – asphalt (91% of buildings), metal (6.9%), rubber membrane (0.77%) and tar (0.94%) and their emissivities were calculated. The building vectors were overlaid on the atmospherically corrected radiance image and using (9), black body radiant emittance for those pixels was calculated. By inverting

Fig. 3. Rooftop temperature error charts for asphalt and metal roofs. The model estimated the temperature of majority asphalt roofs within 1°C; however the temperature of flat metal roofs was highly underestimated.

Planck's law kinetic temperature of building roofs were computed.

### III. RESULTS AND DISCUSSION

To validate the model, kinetic temperature of 22 rooftops measured during the field run were compared with the temperature obtained from the images. These validation targets were not a part of the regression model. Due to the overlaps in the flight line, some houses had multiple images, thus a total of 47 data points were used to validate. The Root Mean Square Error (RMSE) for asphalt roofs was 0.748°C and for metal roofs was 11.816°C. Figure 3 shows the error distribution and it can be observed that the error appears random for asphalt roofs but it is negative for metal roofs meaning the model has grossly under-estimated the temperature of all metal roofs. This indicates the radiant behavior of flat metal roofs is clearly not similar to angled asphalt roofs and more research needs to be done on estimating the temperature of metal roofs using this method. It is to be noted that in this study all the calibration field targets used for ELC regression were impervious targets made of

TABLE 2

| Case | Image number | Flight line | Average roof temperature |
|------|--------------|-------------|--------------------------|
| 1    | 3749         | 39          | 5.85860                  |
| | 3750 | 39 | 5.65984 |
| | 3029 | 32 | -5.82849 |
| 2 | 3030 | 32 | -6.01958 |
| | 3557 | 37 | -5.80157 |
| 3 | 1440 | 18 | -6.19326 |
| | 1441 | 18 | -6.69247 |

Temperature error due to wind gusts.

asphalt and cement and none of metal. One suggestion would be to treat metal roofs as a different population and derive new ELC regression parameters for them.

The ambient ground conditions during the study were -1.49°C air temperature, -3.84°C dew point, 84.08% humidity, 6.75 m/s wind speed and a wind chill of -4.88°C. However occasional gusts in the order of 9.38 m/s disrupted the ground temperature measurement by reducing the temperature with convective cooling. This is particularly evident while comparing the roof temperature in overlap images and is shown in Table 2. Case 1 in the table represents two consecutive images along the same flight line which were taken a few seconds apart. The 0.2°C difference in temperature could be due to micro-climate such as wind gusts, changing angle of observation, change in atmospheric condition and instrumental variation. However, we suspect the microclimate to have a dominant a role. In case 2, there is a difference in temperature along the same flight line, however in the adjacent flight line, the temperature comes back to the initial condition representing the sudden and short nature of these gusts. Case 3 represents one of the worse cases. Thus the slope of the ELC model represents atmospheric absorption, reduction in the number of photons reaching the sensor while the offset represents atmospheric contribution in the form of re-radiation and scattering. Other assumptions made in this study were considering all roofs as flat – neglecting anisotropic nature of emissivity. In the field measurements, while converting radiant temperature to kinetic temperatures, the contribution of the few feet of air was neglected, in particular the absorption effects by oxygen and ozone was neglected.

The temperature map of building roofs shown in Figure 4 would be subjected to GIS analysis to determine poorly performing roofs, hot spots within a roof and to quantitatively categorize buildings based on their heat loss. The resulting map is intended to be hosted as a web mapping application for public access. A software library in Python programming language was created to perform emissivity calculations, inter conversion between radiance & temperature and atmospheric & emissivity corrections. Numerical integration using trapezoidal formula was adopted in place of integration. In the place of inversion of Planck's law a look-up table was created linking band radiance with temperature.

## IV. CONCLUSION

The problem of extracting kinetic temperature of building rooftops from high spatial resolution aerial thermal images is addressed in this paper. The methodology for collecting and normalizing field temperature data, calculating band emissivity from spectral reflectance curves and atmospheric correction using empirical line calibration technique is presented. For the purposes of identifying buildings with

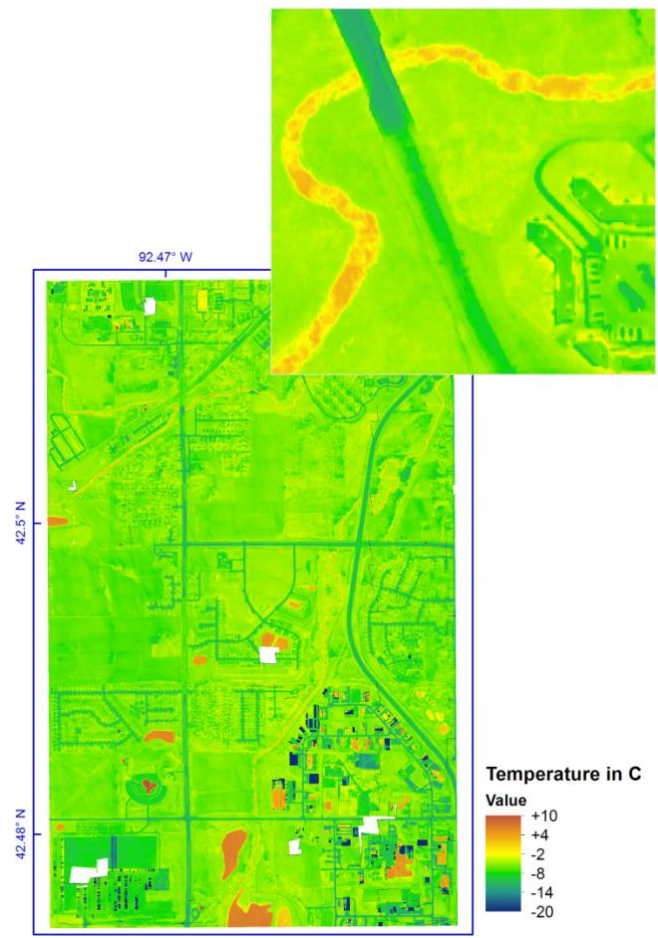

Fig. 4. Thermal map of the study area. Inset shows a zoomed portion of a house, asphalt road, concrete bridge and a creek.

insufficient insulation, the methodology turns out to be sufficiently accurate in estimating rooftop temperature of angled asphalt roofs, while the accuracy of metal roofs is less and needs further analysis

ACKNOWLEDGMENT